\documentclass{aa}

\usepackage{booktabs}
\usepackage{multirow}
\usepackage{txfonts,graphicx} 
\usepackage[colorlinks=true, allcolors=blue, citecolor=blue, linkcolor=blue, urlcolor=blue]{hyperref}
\usepackage[position=top]{caption}
\usepackage[normalem]{ulem}

\begin{document}

   \title{Semi-supervised classification of stars, galaxies and quasars using K-means and random-forest approaches}

   \author{V. Asadi\inst{1}~\href{https://orcid.org/0009-0005-8897-2385}{\includegraphics[height=0.8em]{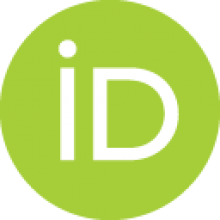}} \and H. Haghi\inst{1,2,3}~\href{https://orcid.org/0000-0002-9058-9677}{\includegraphics[height=0.8em]{orcid-ID.png}} \and A.H. Zonoozi\inst{1,2}~\href{https://orcid.org/0000-0002-0322-9957}{\includegraphics[height=0.8em]{orcid-ID.png}}}

   \institute{Department of Physics, Institute for Advanced Studies in Basic Sciences (IASBS), PO Box 11365-9161, Zanjan, Iran\\
              \email{vahidasadi@iasbs.ac.ir}
        \and
             Helmholtz-Institut f\"ur Strahlen-und Kernphysik (HISKP), Universit\"at Bonn, Nussallee 14-16, D-53115 Bonn, Germany\\
             \email{haghi@iasbs.ac.ir}
        \and
             School of Astronomy, Institute for Research in Fundamental Sciences (IPM), PO Box 19395 - 5531, Tehran, Iran
             }

   \date{}

  \abstract
   {Classifying stars, galaxies, and quasars is essential for understanding cosmic structure and evolution; however, the vast data from modern surveys make manual classification impractical, while supervised learning methods remain constrained by the scarcity of labeled spectroscopic data.}
   {We aim to develop a scalable, label-efficient method for astronomical classification by leveraging semi-supervised learning (SSL) to overcome the limitations of fully supervised approaches.}
   {We propose a novel SSL framework combining K-means clustering with random forest classification. Our method partitions unlabeled data into 50 clusters, propagates labels from spectroscopically confirmed centroids to 95\% of cluster members, and trains a random forest on the expanded pseudo-labeled dataset. We applied this to the CPz catalog, containing multi-survey photometric and spectroscopic data, and compared performance with a fully supervised random forest.}
   {Our SSL approach achieves F1 scores of 98.8\%, 98.9\%, and 92.0\% for stars, galaxies, and quasars, respectively, closely matching the supervised method with F1 scores of 99.1\%, 99.1\%, and 93.1\%, while outperforming traditional color-cut techniques. The method demonstrates robustness in high-dimensional feature spaces and superior label efficiency compared to prior work.}
   {This work highlights SSL as a scalable solution for astronomical classification when labeled data is limited, though performance may be degraded in lower dimensional settings.}


   \maketitle

\section{Introduction}
The exponential growth of astronomical data from modern sky surveys has made the manual classification of stars, galaxies, and quasars obsolete, necessitating scalable and automated methods \citep[e.g.,][]{ivezic2019lsst}. Traditional classification approaches, such as color-cut or morphological selection \citep[e.g.,] []{dey2019overview,bellstedt2020galaxy} are limited to low-dimensional feature spaces and require manual threshold tuning for each survey. In contrast, machine learning (ML) methods use high-dimensional data, including photometry and morphological features, to achieve superior accuracy, particularly for ambiguous cases \citep[e.g.,][]{clarke2020identifying,cunha2022photometric,cook2024wide}. ML classifiers automatically adapt to survey-specific characteristics, handle noisy or missing data robustly \citep[e.g.,] []{gupta2019dealing,emmanuel2021survey}, and classify stars, galaxies, and quasars simultaneously within a unified framework, avoiding biases inherent to sequential selection rules. Crucially, ML scales seamlessly to upcoming surveys such as the LSST, where manual rule-based methods become impractical \citep[e.g.,][]{ivezic2019lsst}.

Although supervised learning methods have been widely applied for this task \citep[e.g.,] []{cavuoti2014photometric,krakowski2016machine,kurcz2016towards,nakoneczny2019catalog,clarke2020identifying,nakoneczny2021photometric,cunha2022photometric,chaini2023photometric,zeraatgari2024machine,von2024j}, its reliance on large-labeled datasets poses a significant limitation, as spectroscopic confirmations remain costly and scarce. 

Unsupervised learning methods offer an alternative by seeking to find patterns or structures in unlabeled data without prior knowledge of class labels \citep[e.g.,][]{james2023unsupervised,fotopoulou2024review}. Techniques such as clustering (e.g., K-means \citep{macqueen1967some}, HDBSCAN\footnote{Hierarchical density-based spatial clustering of applications with noise} \citep{mcinnes2017hdbscan}), or dimensionality reduction (e.g., PCA\footnote{Principal component analysis} \citep{mackiewicz1993principal}; t-SNE\footnote{t-distributed stochastic neighbor embedding} \citep{Policar2021}; UMAP\footnote{Uniform manifold approximation and projection} \citep{mcinnes2018umap}) can be used to group similar objects or reduce the complexity of the data. For example, \cite{cook2024wide} showcased the effectiveness of unsupervised methods such as UMAP and HDBSCAN for crucial star-galaxy separation. Their approach, applied to photometric data, efficiently classified stars and galaxies with high purity, significantly reducing wasted telescope time by improving target catalogs. Although these methods are powerful for discovering hidden relationships and are not limited by the availability of labeled data, their direct application for classification can be challenging, as the identified clusters may not always directly correspond to known astronomical classes. This often requires post hoc interpretation or additional steps to assign meaningful labels, which can introduce its own set of complexities and potential biases. Additionally, when the characteristics of astronomical objects resemble each other in their boundaries in photometry space, clustering approaches can struggle to merge these distinct populations into a single cluster.

Semi-supervised learning (SSL) offers a promising alternative by leveraging both limited labeled data and abundant unlabeled photometric data \citep[e.g.,][]{reddy2018semi, van2020survey}. SSL employs a variety of strategies, which can be broadly categorized into inductive and transductive approaches. Inductive methods focus on constructing a predictive model capable of generalizing to unseen data. In contrast, transductive methods seek to solve the problem using only the available dataset. The integration of labeled and unlabeled samples is a common practice in astronomy, leading to numerous applications of SSL—ranging from active galactic nucleus (AGN) classification to supernova identification \citep[e.g.,] []{lawlor2016mapping, villar2020superraenn, pantoja2022semi, slijepcevic2022radio}.

As part of this study, we propose a novel SSL framework that combines K-means clustering \citep{macqueen1967some} with random-forest classification \citep{Breiman2001} to classify stars, galaxies, and quasars using only 50 labeled instances—orders of magnitude fewer than required by supervised methods. Our approach first partitions unlabeled data into clusters, propagates labels from spectroscopically confirmed centroids, and then trains a random-forest classifier on the expanded pseudo-labeled dataset. We validate this method on the CPz catalog, demonstrating performance comparable to fully supervised approaches while drastically reducing labeling effort.
The paper is organized as follows: Sect.~\ref{sec:2} describes the CPz dataset, Sect.~\ref{sec:3} outlines preprocessing steps, Sects.~\ref{sec:4} and \ref{sec:5} detail our SSL and supervised methods, Sect.~\ref{sec:6} presents results, and Sect.~\ref{sec:7} discusses broader implications and limitations.
\section{Data}\label{sec:2}\label{sec:2}
To apply our SSL approach to classify stars, galaxies, and quasars, we utilized the CPz catalog developed by \cite{fotopoulou2018cpz} (hereafter FP18). This catalog is designed for classification-aided photometric-redshift (z) estimation and comprises spectroscopically observed sources from multiple surveys, covering a redshift range of 0 to 4. Notably, the Sloan Digital Sky Survey (SDSS) samples dominate at higher redshifts.

The spectroscopic data in the catalog are sourced from several well-known surveys, including SDSS DR12 \citep{alam2015eleventh}, GAMA DR2 \citep{liske2015galaxy}, VIPERS DR1 \citep{garilli2014vimos}, VVDS DR2 \citep{le2013vimos}, PRIMUS DR1 \citep{coil2011prism,cool2013prism}, and 6dF DR3 \citep{jones20046df,jones20096df}. To ensure high data quality, FP18 applied strict filters, retaining only sources with the most reliable spectroscopic redshift measurements. These spectroscopic sources were then cross-matched with photometric detections from various surveys using a narrow angular radius of $1^{\prime\prime}$.

The photometric data in the catalog span a wide range of the electromagnetic spectrum, including optical and infrared wavelengths. Specifically, the catalog incorporates the elements listed below.

\begin{itemize}
    \item Optical filters: u, g, r, i, and z from SDSS DR12 \citep{alam2015eleventh}, CFHTLS-T0007 Wide \citep{hudelot2012vizier}, and KiDS DR2 \citep{de2015first}.

    \item Near-infrared filters: Z, Y, J, H, and $K_{s}$ from the ESO VISTA surveys (VIKING and VIDEO) \citep{edge2013vista,jarvis2013vista}.

    \item Mid-infrared filters: W1 and W2 from the WISE ALLWISE data release \citep{wright2010wide,mainzer2011preliminary,cutri2013explanatory}.
\end{itemize}

\begin{figure}[t]
    \centering
    \includegraphics[width=0.95\linewidth]{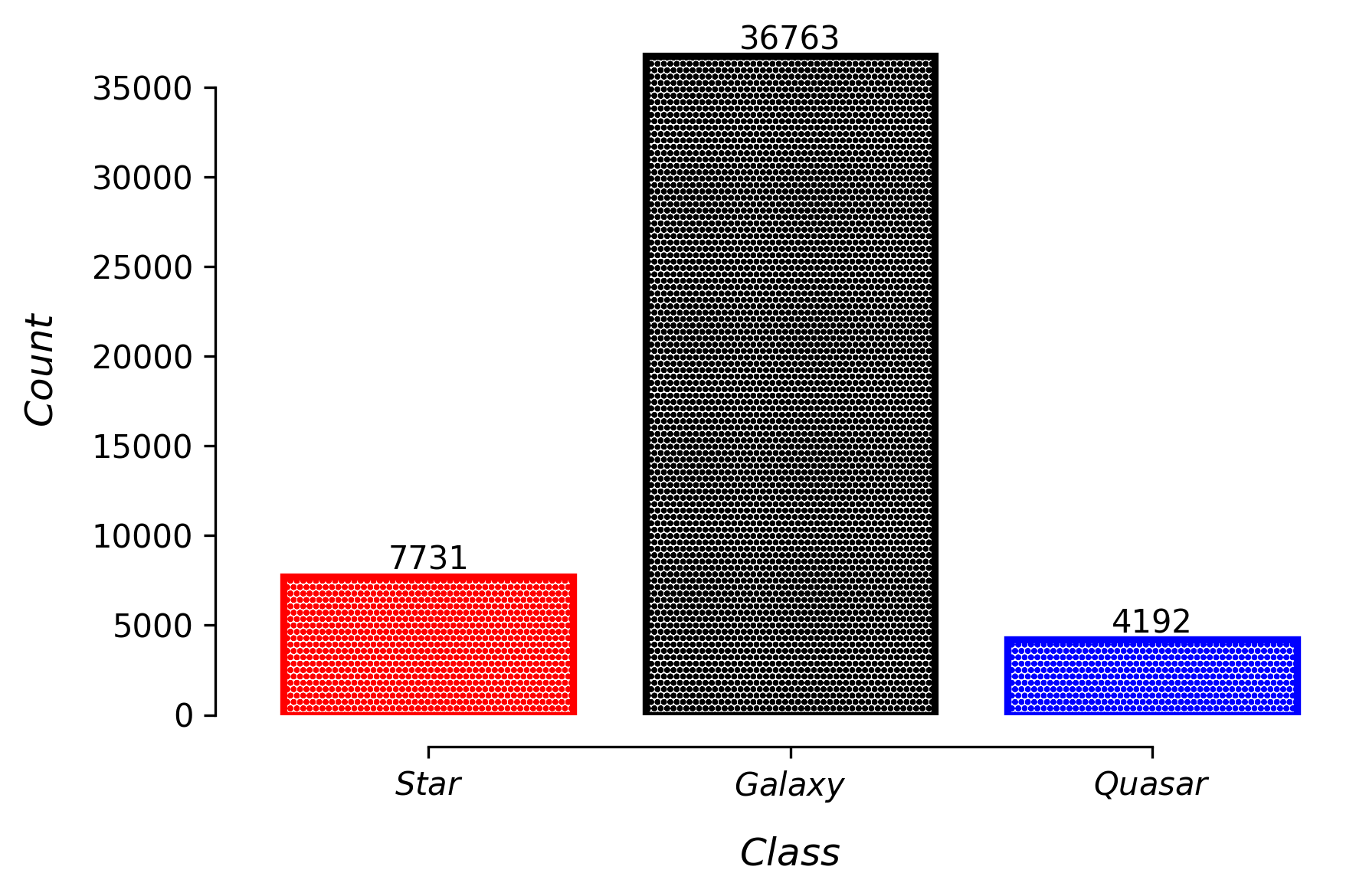}
    \caption{Distribution of sources in FP18 dataset, categorized into stars, galaxies, and quasars. The dataset is imbalanced, with galaxies constituting the majority.}
    \label{fig:fig1}
\end{figure}

\begin{figure*}
    \centering
    \includegraphics[width=0.85\linewidth]{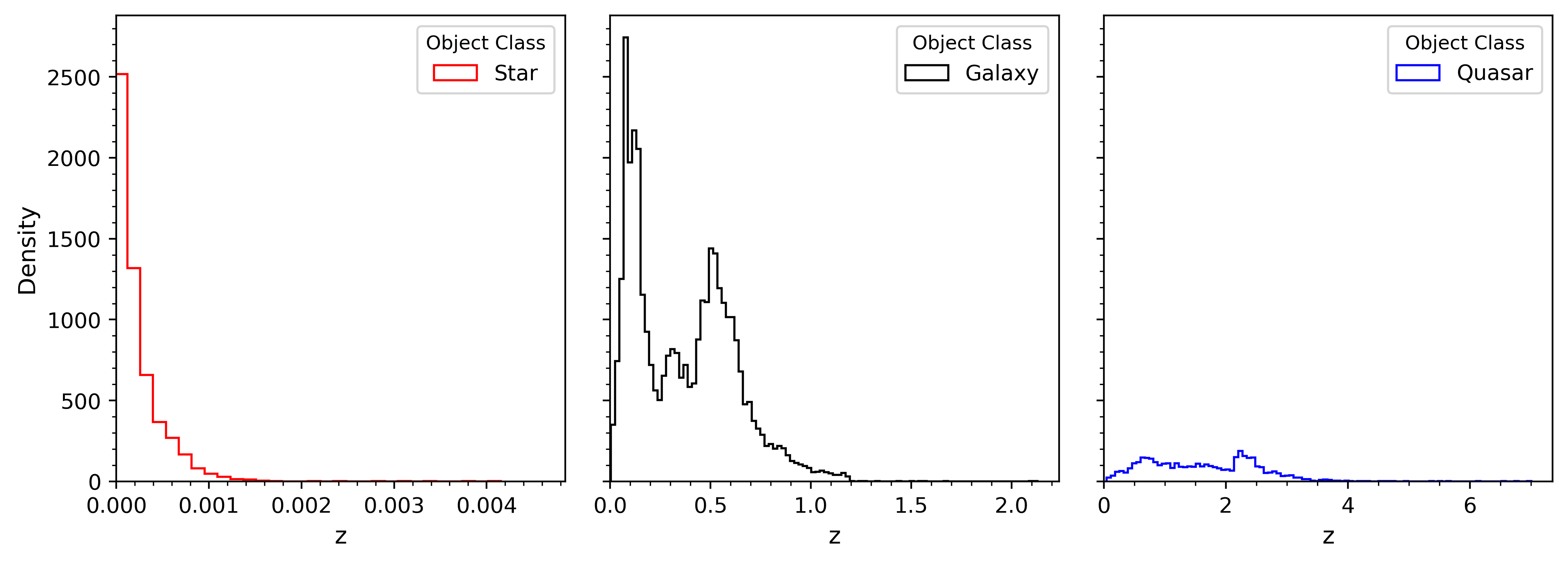}
    \caption{Redshift distribution of spectroscopically confirmed sources in the FP18 catalog, shown by class: stars (red), galaxies (black), and quasars (blue).}
    \label{fig:fig2}
\end{figure*}

\begin{figure}
    \centering
    \includegraphics[width=0.9\linewidth]{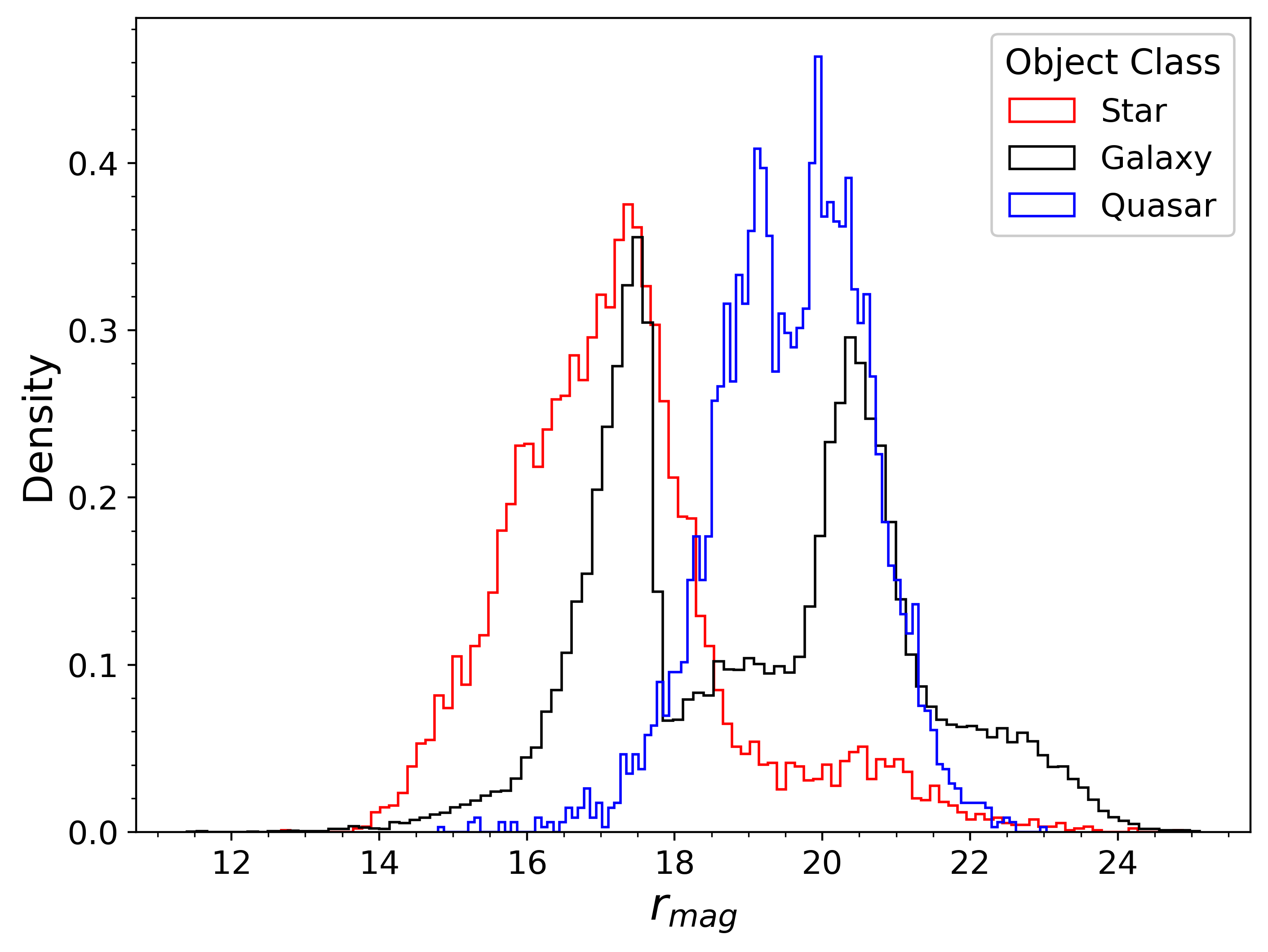}
    \caption{Distribution of $r$-band magnitudes for stars (red), galaxies (black), and quasars (blue).}
    \label{fig:fig22}
\end{figure}

The FP18 dataset consists of 48,686 sources, classified into three categories: stars, galaxies, and quasars. Each source was measured using total and $3^{\prime\prime}$ aperture magnitudes in the optical and near-infrared bands, as well as total magnitudes in the mid-infrared bands. Fig.~\ref{fig:fig1} illustrates the distribution of these classes within the dataset. Notably, FP18 is an imbalanced dataset, with galaxies representing the majority (36,763 out of 48,686 sources).
Figures~\ref{fig:fig2} and~\ref{fig:fig22} present the redshift and r-band magnitude distributions, respectively, for stars, galaxies, and quasars in our sample.

\section{Pre-processing}\label{sec:3}
\subsection{Feature engineering}
The FP18 catalog provides complete photometric measurements with no missing values across all bands, ensuring full feature availability for our analysis. For training our ML classifiers, we utilized the unique color combinations derived from both the total and $3^{\prime\prime}$ aperture magnitudes (e.g., u-g, g-$K_{s}$), resulting in a total of 190 features. 

To ensure consistent feature scaling and improve the learning process for our models, we standardized all datasets using the \texttt{StandardScaler} function from the scikit-learn library. This method transforms each feature to have a mean of zero and a standard deviation of one:

\begin{equation}
X_{\text{scaled}} = \frac{X - \mu}{\sigma}
\end{equation}

where $\mu$ is the mean and $\sigma$ is the standard deviation of each color feature.

\subsection{Data splitting}\label{sec:3.2}
The FP18 dataset was partitioned into training (80\%) and testing (20\%) sets using stratified random sampling. This approach preserved the original class distributions (stars, galaxies, and quasars) in both subsets, ensuring representative sampling across all categories while mitigating potential biases in redshift and spatial distributions. The resulting training set comprised 38,948 labeled instances, with 9,738 instances reserved for testing.

\begin{figure*}[t]
    \centering
    \includegraphics[width=\linewidth]{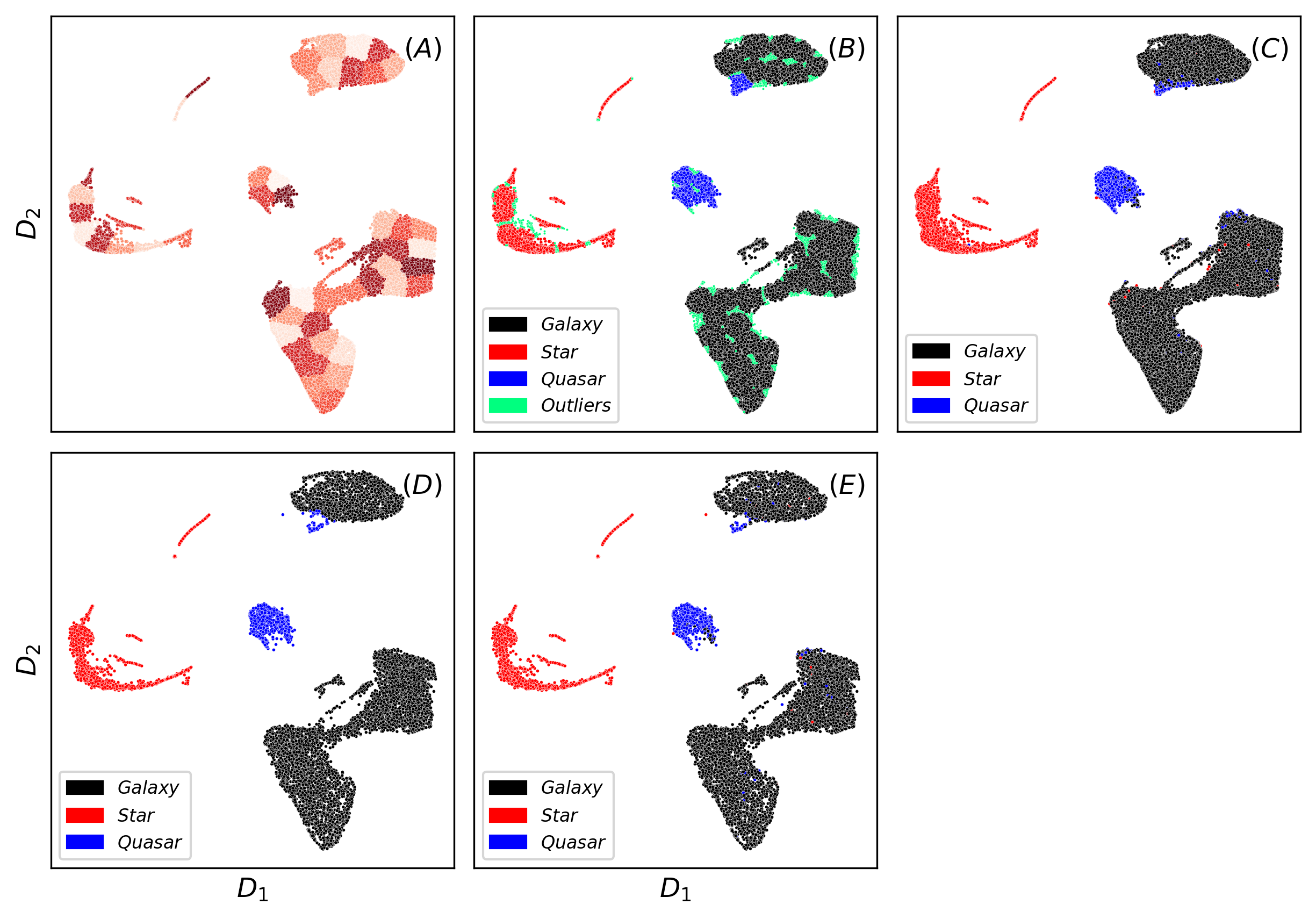}
    \caption{(A) UMAP projection of FP18 training set partitioned into 50 clusters via K-means. (B) Label propagation from spectroscopically confirmed centroids to 95\% of cluster members. (C) Ground-truth spectroscopic labels for FP18 training set. (D) Random forest predictions on FP18 testing set. (E) Ground-truth spectroscopic labels for FP18 testing set. Axes D1 and D2 represent the reduced dimensions and do not carry physical meaning. The qualitative similarity between panels (D) and (E) demonstrates the effectiveness of our SSL approach.}
    \label{fig:fig3}
\end{figure*}

\section{Semi-supervised classification}\label{sec:4}
Semi-supervised learning is a machine-learning approach that leverages both a small amount of labeled data and a large amount of unlabeled data. This method is particularly useful in scenarios where labeled data are scarce or costly to obtain, as it combines the strengths of supervised and unsupervised learning to improve model performance. While various techniques exist \citep[e.g.,][]{reddy2018semi,van2020survey}, we focused on clustering-based SSL. Before detailing our specific SSL method, we briefly introduce K-means \citep{macqueen1967some} and random-forest \citep{Breiman2001} methods, the two algorithms central to our approach.

\subsection{K-means}
K-means is an unsupervised clustering algorithm that partitions a dataset into a predefined number of clusters (K). The goal of the algorithm is to group similar data points together while maximizing the separation between clusters. The process begins by randomly initializing the centroids for each cluster. Data points are then assigned to the nearest centroid, forming initial clusters. The centroids are subsequently recalculated as the mean of their assigned data points. This iterative process continues until the centroids converge, typically determined by a predefined threshold or a maximum number of iterations. Common distance metrics used in K-means include Euclidean and Manhattan distances. Despite its simplicity and efficiency, K-means assumes that clusters are spherical and similar in size, which may not hold true for all datasets.

\subsection{Random-forest classifier}
Random forest is an ensemble learning method that combines multiple decision trees. It constructs a "forest" of trees by training each one on a random subset of the original data (bootstrap sampling) and a random subset of features (feature bagging). During prediction, each tree casts a vote, and the final prediction is determined by majority voting. This approach reduces overfitting and enhances generalization performance. The random-forest method is particularly well-suited for our task due to its ability to handle high-dimensional data and mitigate overfitting through ensemble learning.

\subsection{Model construction}
Our dataset labeling process involves two steps:
\begin{itemize}
    \item Step 1: K-means clustering and label propagation 

     We partitioned the training set into 50 clusters using the K-means algorithm\footnote{\url{https://scikit-learn.org}}. For each cluster, data points closest to the centroid were assigned labels (star, galaxy, or quasar) based on spectroscopic classifications. These labels were then propagated to the 95\% of instances nearest to the centroid within each cluster (by Euclidean distance), with the outermost 5\% classified as outliers. The 5\% threshold was selected to minimize potential overlap or misclassification at the boundaries between stars, galaxies, and quasars. Our sensitivity analysis across 90-99\% propagation thresholds showed minimal impact on classification performance (stars: $\sigma_{F1-score}$\footnote{F1 score standard deviation} = 0.0002; galaxies: $\sigma_{F1-score}$ = 0.0002; quasars: $\sigma_{F1-score}$ = 0.0021; Eq.~\ref{eq:4}), justifying our selection of 95\% as the optimal middle ground.

     Although stars, galaxies, and quasars generally occupy distinct regions in high-dimensional color space, transitional or rare subpopulations (e.g., low-redshift quasars and compact galaxies) may still reside in proximity of one another. Additionally, K-means’ assumption of spherical clusters does not perfectly align with the intrinsic geometry of our data. A small K value risked merging distinct subpopulations (e.g., misclassifying quasars as part of galaxy clusters). To mitigate these risks, we selected K = 50. While this number is somewhat arbitrary, it ensured that we avoided the aforementioned issues of cluster overlap and geometric mismatch. Our sensitivity analysis (Sect.~\ref{sec:6.3}) demonstrates that K = 50 provides optimal performance, balancing cluster granularity against computational efficiency while avoiding issues of cluster overlap and geometric mismatch.

    \item Step 2: Random-forest classification

     We trained a \texttt{RandomForestClassifier}\footnote{\url{https://scikit-learn.org}} on the pseudo-labeled dataset from Step 1, using default parameters (100 trees, Gini impurity, unlimited max depth) as preliminary hyperparameter tests showed diminishing returns beyond this configuration. The trained model was then used to predict labels for the testing set. Finally, the predicted labels were compared to the spectroscopically confirmed true labels using evaluation metrics outlined in Sect.~\ref{sec:61}.
    
\end{itemize}

To visualize this approach, we applied unsupervised UMAP\footnote{\url{https://pypi.org/project/umap-learn}} \citep{mcinnes2018umap} to the FP18 training set, reducing it to two dimensions, followed by K-means clustering into 50 clusters. Fig.~\ref{fig:fig3} illustrates the SSL classification workflow. Panel (A) displays the 50 clusters identified in the FP18 training set after UMAP dimensionality reduction; panel (B) demonstrates label propagation, where labels from spectroscopically confirmed centroids (stars, galaxies, quasars) are extended to 95\% of cluster members, with the outermost 5\% marked as outliers; panel (C) provides the ground-truth spectroscopic labels for the FP18 training set; panel (D) shows random-forest model predictions on the FP18 testing set, trained on the labels propagated in step 1; and panel (E) provides the ground-truth spectroscopic labels for the FP18 testing set. 

Comparing panels (B) and (C) reveals high label-propagation accuracy, with average purity of 97\% across classes (stars: 99\%; galaxies: 99\%; quasars: 93\%). The strong visual agreement between predicted (D) and actual (E) distributions confirms our method's effectiveness in reproducing the true astronomical classifications. Minor discrepancies occur primarily at class boundaries where photometric properties overlap.

To address the instability of K-means initialization arising from the algorithm's sensitivity to the initial placement of centroids \citep[e.g., see][]{celebi2013comparative}, we repeated the labeling process 100 times. We then aggregated the predictions from all 100 runs using majority voting. This ensemble approach ensures that the final classification reflects the most consistent outcome, thereby minimizing the influence of outlier predictions caused by suboptimal initializations.

\section{Supervised classification}\label{sec:5}
To establish a rigorous performance benchmark, we implemented a \texttt{RandomForestClassifier} using the complete FP18 labeled training set (38,948 instances), maintaining identical training and test splits as our SSL approach (random seed = 42; see Sect.~\ref{sec:3.2}). Initial evaluation with default parameters (100 trees, unlimited maximum depth, 1 minimum sample per leaf) demonstrated that further optimization provided diminishing returns, as confirmed through stratified random 10\% hold-out validation.

We conducted random-search cross-validation (50 iterations) across the hyperparameter space, varying the number of trees (50--500), maximum depth (5--50), and minimum samples per leaf (1--20). This exhaustive search yielded only marginal improvements ($\leq$0.2\% in F1 score; Eq.~\ref{eq:4}).

Additionally, we observed similarly negligible improvements ($<$0.3\% F1 score increase) for minority classes (stars and quasars) when applying class imbalance strategies, including class weighting and SMOTE\footnote{Synthetic minority ver-sampling technique.} oversampling \citep{chawla2002smote}.

\section{Results}\label{sec:6}
\subsection{Performance metrics}\label{sec:61}
To evaluate and compare the performance of our classifiers, we utilized the confusion matrix \citep[e.g., see ] []{stehman1997selecting} along with derived metrics such as precision (purity), recall (completeness), and F1 score \citep[e.g., see ] []{christen2023review}. These metrics provide a comprehensive understanding of how well our classifiers distinguish stars, galaxies, and quasars, especially in the context of our imbalanced dataset where accuracy can be misleading.

The confusion matrix is a table used to describe the performance of a classification model on a set of test data for which the true values are known. For a multi-class classification problem with three classes (stars, galaxies, and quasars), the confusion matrix includes the following terms:

\begin{itemize}
\item True positive (TP\textsubscript{i}): the number of correctly predicted observations for class $i$
\item False positive (FP\textsubscript{i}): the number of observations incorrectly predicted as class $i$
\item False negatives (FN\textsubscript{i}): the number of observations of class $i$ incorrectly predicted as other classes
\end{itemize}

Table~\ref{tab:tab0} illustrates a confusion matrix for a three-class classification.
Precision for class $i$ is the ratio of correctly predicted observations for class $i$ to the total predicted observations for class $i$. It measures the accuracy of positive predictions for class $i$:

\begin{equation}
\text{Precision}_i = \frac{TP_i}{TP_i + FP_i}
\label{eq:2}
.\end{equation}

Recall for class $i$ is the ratio of correctly predicted observations for class $i$ to all actual observations of class $i$. It measures how well the model captures instances of class $i$:

\begin{equation}
\text{Recall}_i = \frac{TP_i}{TP_i + FN_i}
\label{eq:3}
.\end{equation}The F1 score for class $i$ is the harmonic mean of precision and recall for class $i$, providing a balance between the two. It is particularly useful when the class distribution is imbalanced:

\begin{equation}
\text{F1 Score}_i = 2 \times \frac{\text{Precision}_i \times \text{Recall}_i}{\text{Precision}_i + \text{Recall}_i}
\label{eq:4}
.\end{equation}These metrics together provide a comprehensive evaluation of the classification performance, helping us to identify the most effective method.

\begin{table}
    \centering
    \caption{Three-class confusion matrix.}
    \label{tab:tab0}
    \begin{tabular}{c|ccc}
        \hline
        & \multicolumn{3}{c}{Predicted} \\
        \cmidrule{2-4}
        Actual & Class 1 & Class 2 & Class 3 \\
        \midrule
        Class 1 & TP\textsubscript{1} & FP\textsubscript{1,2} & FP\textsubscript{1,3} \\
        Class 2 & FP\textsubscript{2,1} & TP\textsubscript{2} & FP\textsubscript{2,3} \\
        Class 3 & FP\textsubscript{3,1} & FP\textsubscript{3,2} & TP\textsubscript{3} \\
        \hline
    \end{tabular}
    
    \smallskip
    \footnotesize
    \raggedright\textbf{Notes.} Rows represent actual classes, and columns represent predicted classes. TP\textsubscript{i} denotes true positives for class \(i\), and FP\textsubscript{i,j} denotes false positives where class \(i\) is predicted as class \(j\).
\end{table}

\begin{table*}
    \caption{Performance comparison of SSL and supervised learning.}
    \centering
    \begin{tabular}{c|cc|c|c|c}
        \hline
        \hline
        Kind & Metric & N\_Labeled samples & Precision & Recall & F1 score \\
        \midrule
                             & Star    & 50     & $0.995$ & $0.982$ & $0.988$ \\
        SSL & Galaxy  & 50     & $0.985$ & $0.993$ & $0.989$ \\
                             & Quasar  & 50     & $0.941$ & $0.899$ & $0.920$ \\       
        \midrule
                             & Star    & 38,948 & $0.995$ & $0.986$ & $0.991$ \\
        Supervised learning      & Galaxy  & 38,948 & $0.987$ & $0.994$ & $0.991$ \\
                             & Quasar  & 38,948 & $0.953$ & $0.909$ & $0.931$ \\
        \hline
    \end{tabular}
    
    \smallskip
    \label{tab:tab1}
\end{table*}

\begin{figure*}
    \centering
    \includegraphics[width=0.65\linewidth]{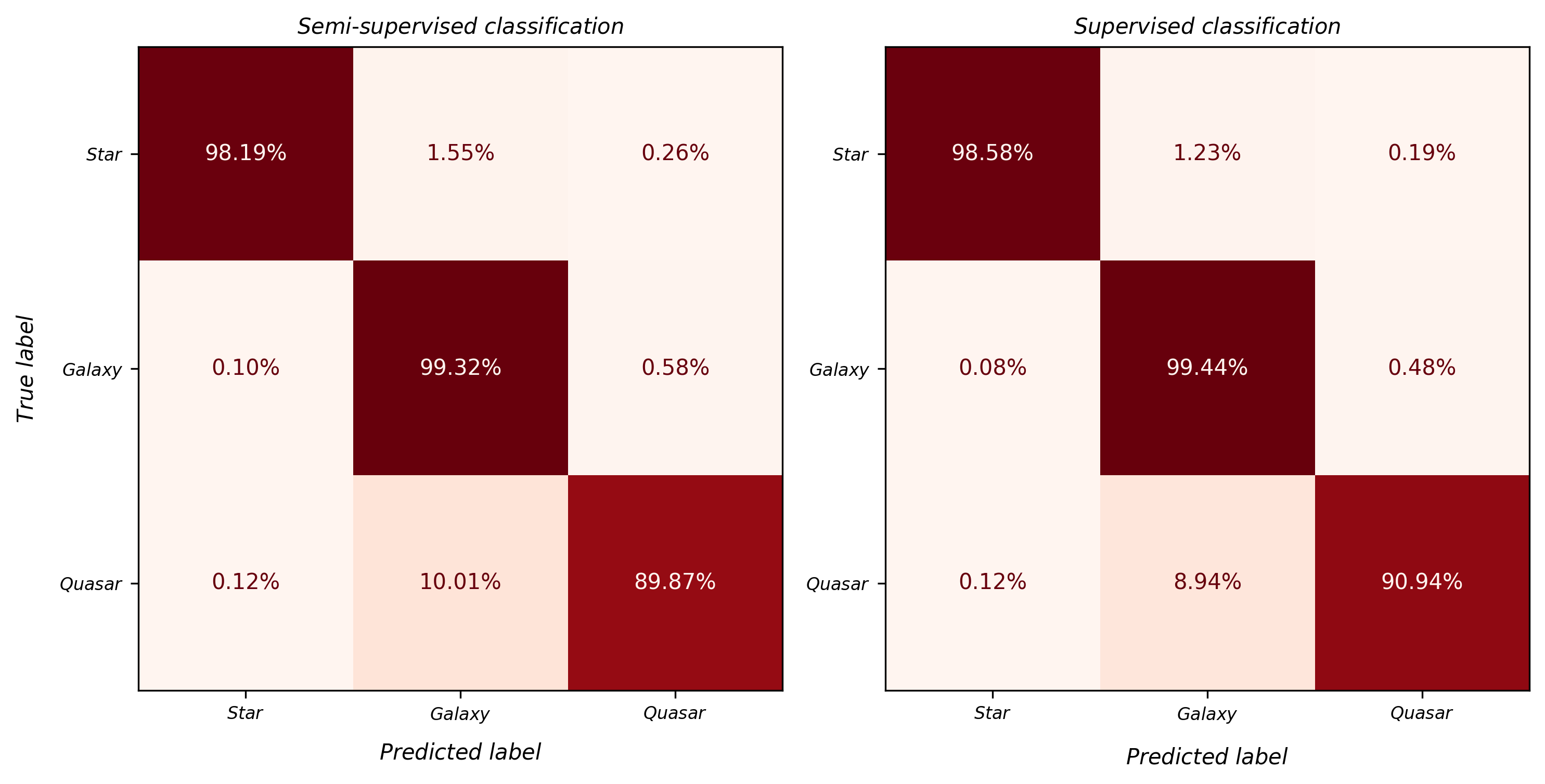}
    \caption{Confusion matrices for supervised and SSL classifiers on the FP18 dataset. The matrices illustrate the classification performance for stars, galaxies, and quasars.}
    \label{fig:fig4}
\end{figure*}

\subsection{Performance of classifiers}\label{sec:6.2}
Table~\ref{tab:tab1} compares the performance of SSL and supervised classification methods on the FP18 dataset, which consists of 190 features, as detailed in Sect.~\ref{sec:3}. The SSL approach utilized only 50 labeled samples, strategically selected using K-means clustering (see Sect.~\ref{sec:4}), whereas the supervised method was trained on the entire training dataset containing 38,948 labeled samples. To evaluate the effectiveness of the models in classifying stars, galaxies, and quasars in the testing dataset, we employed precision (purity), recall (completeness), and F1 score metrics. This comparison demonstrates that the SSL achieves a performance level close to that of supervised learning for stars and galaxies, with a slight performance gap for quasars.

The F1 scores for stars and galaxies were 98.8\% and 98.9\% for the SSL method, and 99.1\% and 99.1\% for the supervised method, respectively. For quasars, the supervised classifier showed a slight advantage, with an F1 score of 93.1\% compared to 92.0\% for the SSL approach. This one-percentage-point difference stemmed from marginal gaps in both precision (supervised: 95.3\%; SSL: 94.1\%) and recall (supervised: 90.9\%; SSL: 89.9\%).

\begin{figure}
    \centering
    \includegraphics[width=0.9\linewidth]{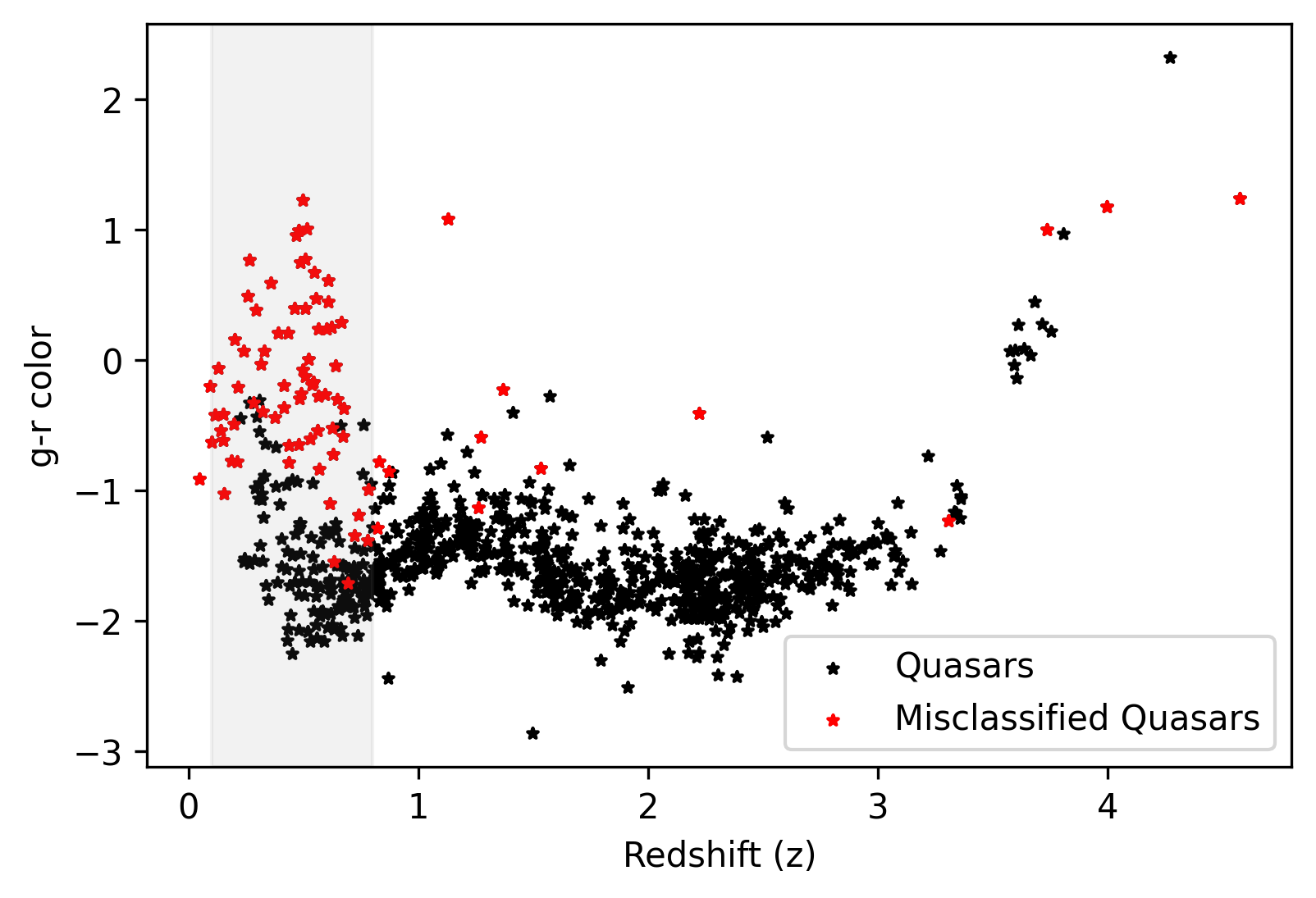}
    \caption{Misclassified quasars (red) versus correctly classified quasars (black) in redshift (z) and $g - r$ color space for the SSL classifier. Most misclassifications occur at z<1, where host galaxy light dilutes AGN signatures, making quasars photometrically resemble galaxies.}
    \label{fig:fignew}
\end{figure}

The observed differences in F1 scores, particularly for quasars, prompted a closer examination of the misclassification patterns. By analyzing the confusion matrices presented in Fig.~\ref{fig:fig4}, we identified that misclassification was negligible for galaxies across both classifiers. For stars, misclassification primarily involved galaxies, with rates of 1.2\% for the supervised method and 1.5\% for the SSL method. For quasars, misclassification also predominantly involved galaxies, suggesting overlapping spectral features between these two classes. The misclassification rates for quasars were 8.9\% for supervised learning and 10.0\% for SSL. For both classifiers, these misclassified quasars mainly related to z<1. As shown in Fig.~\ref{fig:fignew} for the SSL classifier, approximately 90\% of quasar misclassifications occurred within this redshift range. This trend aligns with known astrophysical properties of quasars; at lower redshifts, the host galaxy’s light becomes more prominent relative to the AGNs in photometric data, causing quasars to resemble galaxies \citep[e.g.,  ] []{richards2001colors,stern2012mid}. Additionally, low-redshift quasars often exhibit weaker broad emission lines due to lower accretion rates or obscuration, further blurring the distinction from galaxies \citep[e.g.,  ] []{hewett2010improved}.

\subsection{Sensitivity analysis of K-means clustering}\label{sec:6.3}
\begin{figure}
    \centering
    \includegraphics[width=0.85\linewidth]{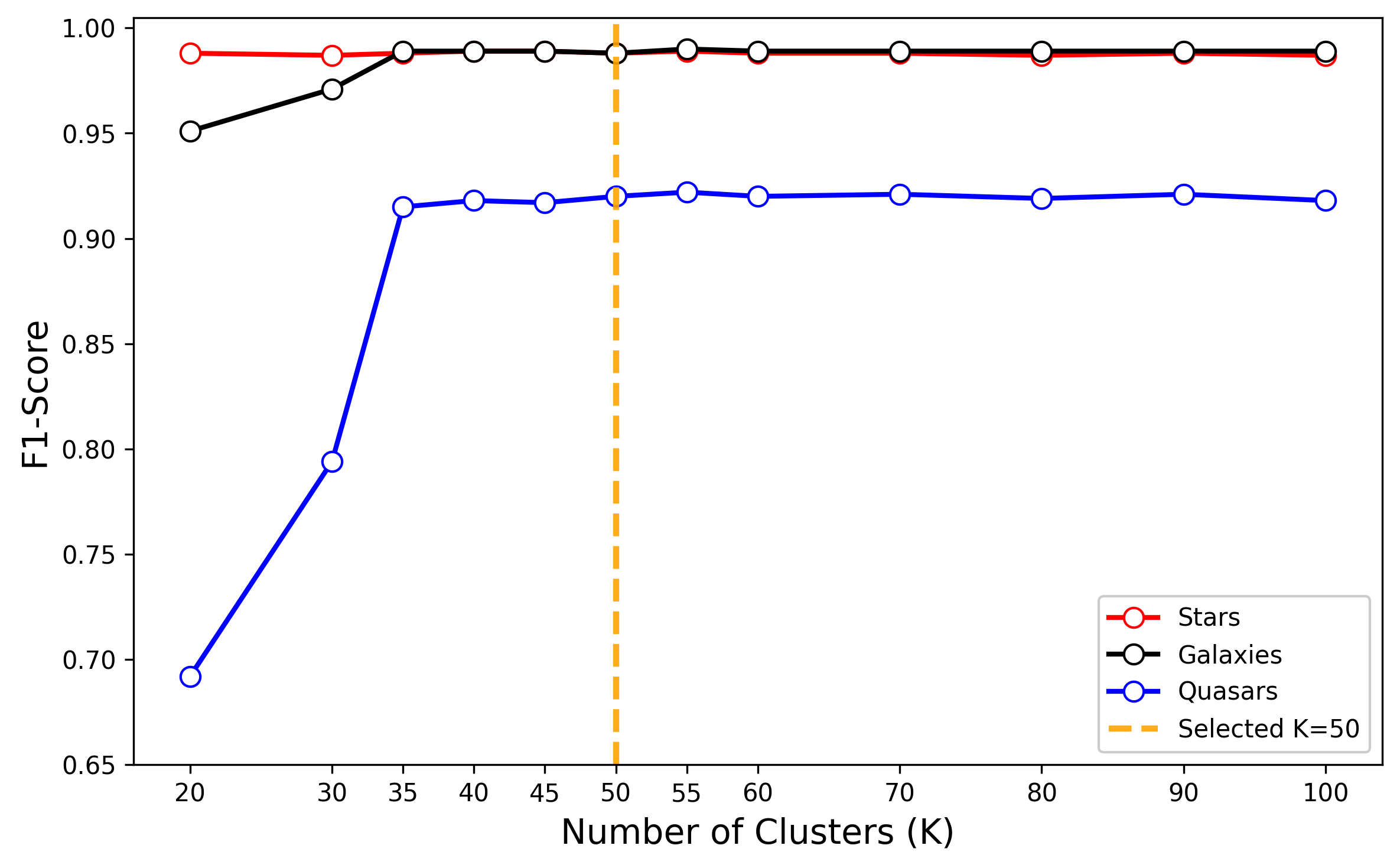}
    \caption{F1 scores remain stable for K$\ge35$: galaxies (black squares, $0.989\pm 0.001$), stars (red circles, $0.988\pm0.001$), and quasars (blue triangles, $0.920\pm0.003$). The dashed orange line marks our chosen K=50.}
    \label{fig:fig00}
\end{figure}

To assess the robustness of our SSL method to the choice of K in K-means clustering, we evaluated F1 scores across K=20-100 (Fig.~\ref{fig:fig00}). The results demonstrate strong stability for stars (F1 score $>0.980$ for all K$>20$) and galaxies (F1 score rising sharply from 0.950 at K=20 to 0.989 at K=30, and then maintaining a $>0.980$ performance level). Quasars show greater initial sensitivity, with the F1 score improving from 0.690 at K=20 to 0.915 at K=35 as finer clusters better capture their diversity, then stabilizing at $0.920\pm0.003$ for K$>35$.

The method shows consistent performance across $K\in[30,100]$, with F1 variations of $\le1\%$ for stars, galaxies and $\le0.3\%$ for quasars. We selected K=50 as an optimal balance between cluster granularity and computational efficiency, avoiding both under-clustering (K$<30$, where galaxy and quasar performance degrades) and over-segmentation (K$>70$, which offers no accuracy gains). This stability confirms that our approach is not overly sensitive to the exact choice of K within this range.

\section{Discussion}\label{sec:7}
Our study demonstrates that a SSL framework combining K-means clustering with random-forest classification achieves performance levels comparable to those of fully supervised methods for classifying stars, galaxies, and quasars, even when using only a small number of labeled instances (as detailed in Sect.~\ref{sec:4}). This approach effectively addresses the challenge of limited labeled data, where manual annotation is often costly and time-consuming.

For this study, we utilized the dataset from \cite{fotopoulou2018cpz} and constructed 190 unique colors from its photometry data. We then split the dataset into training and testing sets to enable a comparison between our SSL method and a random-forest supervised method, which served as a benchmark. The F1 scores achieved by the SSL method using 50 labeled data points were 98.8\%, 98.9\%, and 92.0\% for stars, galaxies, and quasars, respectively. In comparison, the supervised method achieved F1 scores of 99.1\%, 99.1\%, and 93.1\% for the same classes using 38,948 labeled data points.

\begin{figure*}
    \centering
    \includegraphics[width=0.85\linewidth]{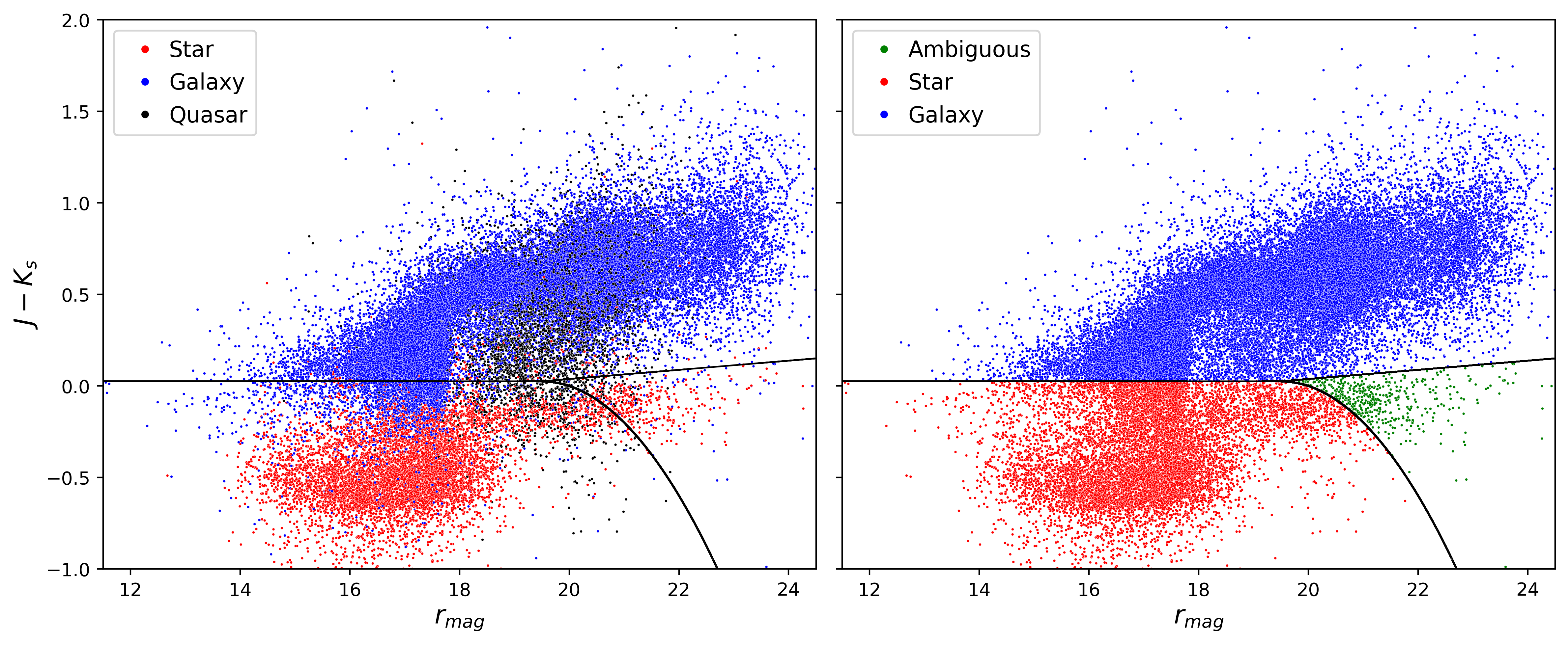}
    \caption{Classification of astronomical sources in (J-$K_{s}$) versus r-band magnitude space. Left: Spectroscopic classification with galaxies (blue), stars (red), and quasars (black). Right: Color-cut classification distinguishing stars (red), galaxies (blue), and ambiguous sources (green). The solid lines mark the empirical boundaries for star-galaxy separation (Eq.~\ref{eq:5}).}
    \label{fig:fig6}
\end{figure*}

\subsection{Impact of morphological features on classification}
Our study focused on developing a purely photometric classifier to maximize applicability across datasets lacking high-resolution morphology. While color features alone achieved strong performance (achieving quasar F1 scores of 0.920 for SSL and 0.931 for supervised classification), incorporating half-light radius (HLR) measurements from the FP18 dataset (Y, H, J, and K bands) further refined the classification. By including HLR values and their ratios and filtering missing and unrealistic values ($0'' < HLR < 20''$), we retained 43,348 high-confidence sources.

The addition of morphological features notably improved quasar identification, with F1 scores increasing to 0.933 (SSL) and 0.948 (supervised) --- representing relative improvements of 1.4\% and 1.8\%, respectively. This enhancement demonstrates that even limited morphological data, when combined with photometry, can improve quasar separation, likely by reducing contamination from compact galaxies or stars. However, the marginal gains for stars and galaxies (F1 score improvements $\leq$0.2\%) suggest that color remains the primary distinguishing factor for these classes.

\subsection{Comparison with prior work}
Our SSL method demonstrates competitive performance compared to previous studies on the same dataset. \cite{lourens2024supervised} employed sharpened dimensionality reduction (SDR), a technique that preconditions high-dimensional data to sharpen clusters and improve separation in 2D projections. They combined SDR with a deep neural network (SDR-NNP) to address scalability and out-of-sample projection limitations. The resulting 2D projections, which preserve neighborhood and distance relationships, were used to train a simple k-nearest-neighbors (k-NN) classifier. This approach achieved F1 scores of 98.7\%, 99.1\%, and 92.3\% for stars, galaxies, and quasars, respectively. In another study, \cite{logan2020unsupervised} used HDBSCAN. They optimized hyperparameters and input attributes for three separate HDBSCAN runs, each targeting a specific object class, and consolidated the results to achieve final classifications optimized for F1 scores. Pre-processing steps included feature selection and dimensionality reduction using random-forest and PCA methods. Their method achieved F1 scores of 98.6\%, 98.7\%, and 91.1\% for stars, galaxies, and quasars, respectively. 

The key advantage of our method over these studies is its simplicity and minimal reliance on labeled data (only 50 instances). Additionally, our approach does not require complex preprocessing steps, making it more efficient and easier to implement.

\subsection{Unsupervised benchmark: UMAP + HDBSCAN performance}
Unsupervised classification methods provide a label-free alternative for astronomical object classification. The combined UMAP+HDBSCAN approach \citep{cook2024wide} has shown superior performance in star-galaxy separation compared to traditional methods. Following their methodology, we implemented this unsupervised pipeline to establish a performance benchmark for our SSL technique, allowing for direct comparison between the label-dependent and label-free classification paradigms.

Our pipeline first reduces the 190-dimensional color feature space to ten dimensions using UMAP (\texttt{number\_of\_neighbors}=200), and then it performs clustering with HDBSCAN\footnote{\url{https://scikit-learn.org}} \citep{mcinnes2017hdbscan} (\texttt{min\_cluster\_size}=3000, \texttt{min\_samples}=500). This configuration, inspired by \cite{cook2024wide}, effectively eliminates noise assignments (typically labeled as -1 by HDBSCAN) while maintaining clear separation between major classes, achieving the following performances:

\begin{itemize}
\item Stars: precision (purity) = 0.996; recall (completeness) = 0.986
\item Galaxies: precision = 0.980; recall = 0.995
\item Quasars: precision = 0.950; recall = 0.841
\end{itemize}

\begin{figure}[t]
    \centering
    \includegraphics[width=0.8\linewidth]{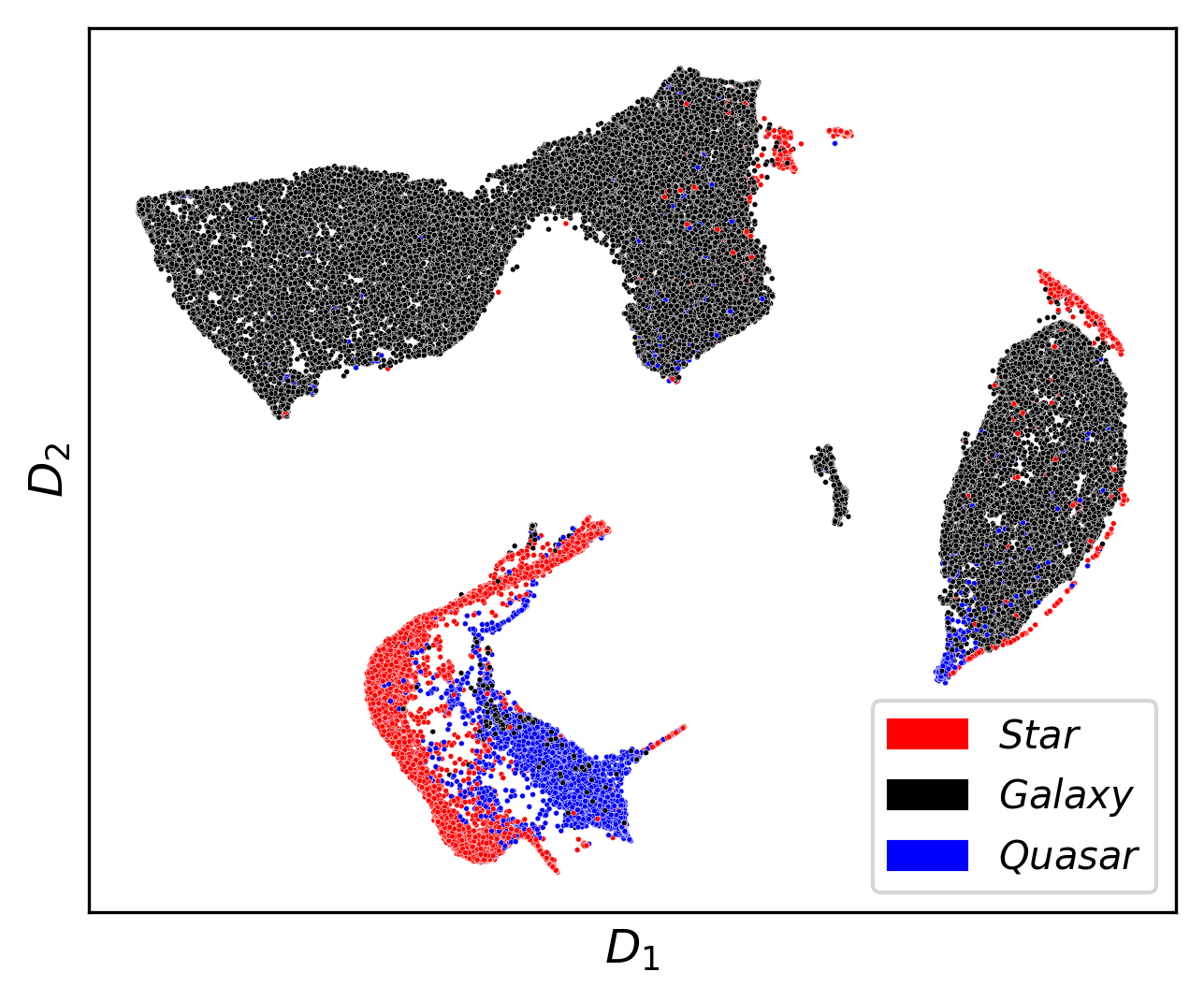}
    \caption{2D UMAP projection of FP18 using 45 colors constructed from optical bands (u, g, r, i, z) and their $3^{\prime\prime}$ aperture counterparts. Axes D1 and D2 represent the reduced dimensions and do not carry physical meaning.}
    \label{fig:fig5}
\end{figure}

The UMAP+HDBSCAN approach demonstrates remarkable effectiveness given its complete lack of labeled training data. For stars and galaxies, it achieves comparable performance to our SSL method (Table~\ref{tab:tab1}), with the density-based clustering successfully capturing these dominant populations.
However, for quasars, we observe worse performance primarily due to less completeness (6\%). This occurs because low-redshift quasars occupy regions that overlap with galaxies (see panel (C) in Fig.~\ref{fig:fig3} and Sect.~\ref{sec:6.2}). Since HDBSCAN identifies clusters based on density connectivity, regions where two classes overlap without a significant density drop are classified as a single cluster. This limitation of the unsupervised approach is further exacerbated when working with reduced-dimensionality feature spaces, where the loss of discriminating features intensifies class overlap (see Fig.~\ref{fig:fig5}). In contrast, the SSL method --which employs K-means clustering and partitions the color space as detailed in Sect.~\ref{sec:4}-- can mitigate this issue by leveraging a small number of labeled data points.

Additionally, after applying HDBSCAN, we identified four distinct clusters. For a quick performance comparison with our SSL method, spectroscopic data guided us in relating each cluster to its respective class. However, a significant challenge with unsupervised learning methods is that, in reality, determining which cluster belongs to which class often requires post hoc interpretation \citep[e.g.,][]{cook2024wide}.

\subsection{Comparison with traditional color-cut methods}\label{sec:7.4}
To contextualize our SSL method's performance against traditional approaches, we compared it with standard color-cut classification in (J-$K_{s}$) versus $r$-band magnitude space (Fig.~\ref{fig:fig6}). The color-cut method employs empirical boundaries \citep{bellstedt2020galaxy} that leverage the tendency of galaxies to have larger (J-$K_{s}$) colors, which increase with magnitude:

\begin{align}
\begin{split}
    (J-K_s) = 0.025, & \qquad\text{if } r < 19.5\\
    (J-K_s) = 0.025+0.025(r-19.5), & \qquad\text{if }r > 19.5\\
    (J-K_s) = 0.025-0.1(r-19.5)^2, & \qquad\text{if }r > 19.5,
\end{split}\label{eq:5}
\end{align}
\\
where sources below the lower boundary are classified as stars, those above the upper boundary are classified as galaxies, and those between the quadratic and linear boundaries for $r\geq19.5$ remain ambiguous. Applying these cuts to the FP18 dataset identified 391 ambiguous sources (0.8\% of the sample).
Compared to the SSL classification results in Table~\ref{tab:tab1}, the color-cut method exhibited systematic biases:

\begin{itemize}
\item Stars: precision (purity) = 0.712 (28\% reduction vs. SSL); recall (completeness) = 0.928 (6\% reduction vs. SSL)
\item Galaxies: precision = 0.902 (6\% reduction vs. SSL); recall = 0.936 (6\% reduction vs. SSL)
\end{itemize}

The superior performance of SSL stems from its ability to leverage high-dimensional color information rather than relying on a single color-magnitude plane. While the (J-$K_{s}$) versus $r$ cuts only utilize two features, our SSL method incorporates a much higher dimensional color-feature space (Sect.~\ref{sec:3}), capturing subtle SEDs that distinguish stars from galaxies. As shown in Fig.~\ref{fig:fig6}, this color-cut method also exhibits significant misclassification of quasars (primarily as galaxies), indicating its inadequacy for quasar classification.

\subsection{Limitations in low-dimensional settings}\label{sec:7.7}
A limitation of our approach is its sensitivity to data dimensionality. As the number of bands --and, consequently, the number of colors-- decreases the performance of our SSL classification method degrades compared to that of the supervised method. This is due to the difficulty in distinguishing stars, galaxies, and quasars in lower dimensional space. In other words, when the dimensionality of the data decreases, the three classes become less separated. Consequently, during the clustering and propagation processes described in Sect.~\ref{sec:4}, some objects may be mislabeled. For example, when using only optical bands (u, g, r, i, and z) and their $3^{\prime\prime}$ aperture counterparts to construct all possible colors (45 colors), the SSL approach achieves F1 scores of 87\%, 98\%, and 81\% for stars, galaxies, and quasars, respectively. This represents a degradation of 12, 1, and 11 percentage points for these objects compared to using all available bands (see Table~\ref{tab:tab1}). Fig.~\ref{fig:fig5} shows the 2D unsupervised UMAP representation of the 45 colors. As can be seen, galaxies are still well separated, but stars and quasars exhibit more overlap compared to the results shown in Fig.~\ref{fig:fig3}, panel D. In contrast, the supervised learning method achieves F1 scores of 97\%, 99\%, and 91\% under the same conditions, with minimal degradation (see Table~\ref{tab:tab1}). However, even in low-dimensional settings, our approach maintains reasonable accuracy while offering the significant advantage of requiring far fewer labeled data points.

In scenarios where the number of bands is limited and the SSL approach exhibits performance degradation due to reduced class separability, active learning \citep{settles2009active} can be employed to enhance the quality of classification. Active learning iteratively selects the most informative unlabeled samples for expert labeling, thereby refining the model's understanding of the decision boundaries in lower dimensional space. By strategically querying labels for objects that lie in ambiguous or overlapping regions of the feature space (e.g., where stars and quasars are less separable), active learning can significantly reduce mislabeling during the clustering and propagation processes. This approach not only mitigates the limitations of SSL in low-dimensional settings, it also maximizes the utility of limited labeled data. Furthermore, the combination of SSL with active learning offers a practical solution for astronomical datasets, where obtaining labeled data is often resource intensive. Future work could explore the integration of active-learning strategies into our framework to further improve classification accuracy while maintaining the efficiency of the SSL method. 

\subsection{Scalability and robustness}
Our SSL approach based on random-forest classification offers advantages for processing data from next-generation astronomical surveys. The method's prediction phase exhibits adaptable scalability characteristics, making it particularly suitable for large-scale photometric surveys such as the LSST and Euclid. Once trained, the computational cost of random-forest predictions depends primarily on the number of trees and their depth rather than the volume of data being processed, enabling the efficient handling of the billions of objects anticipated by the LSST \citep[e.g.,][]{Geurts2006,asadi2025leveraging}. This scalability is further enhanced by the algorithm's inherent parallelizability when implemented on multi-core architectures \citep[e.g.,][]{Breiman2001,Geurts2006,hastie2009elements}.

The random-forest framework demonstrates notable robustness to several practical challenges encountered in survey astronomy. It naturally accommodates missing values during inference since each constituent decision tree makes splits using only available features, with the ensemble's majority voting mechanism effectively mitigating the impact of incomplete data \cite[e,g.,][]{Breiman2001,hastie2009elements}.

Furthermore, the bagging procedure and random feature subsampling provide inherent resistance to moderate noise levels in the input data. While individual trees may be affected by noisy features or outliers, the ensemble collectively dilutes their influence through averaging across many independent predictors. However, systematic noise patterns or extremely low signal-to-noise conditions may degrade performance when they significantly alter the underlying data distributions used for decision tree splits \citep[e.g.,][]{hastie2009elements,zhou2025ensemble}.

\section*{Data Availability}
The FP18 dataset and the implementation code for the SSL classification method are available at the following GitHub repository: \url{https://github.com/vahidoo7/stars-galaxies-quasars-classification}.

\bibliographystyle{aa}
\bibliography{ref}

\end{document}